
\input phyzzx
\def\CR{C_{2R}}
\def\Om{\Omega}
\def\Wl{Wilson loop}
\def\be{\beta}
\def\al{\alpha}
\def\vp{\varphi}
\def\pPi{\partial\Pi}
\def\opi{{1\over\pi}}
\def\muF{\mu_{\rm F}}
\def\tL{\tilde L}

\def\ct#1{\cot{\pi\over\tL}(#1)}
\def\tf{\widetilde f}
\def\tg{\widetilde g}
\def\th{\widetilde h}
\def\bR{\bar R}
\def\Sn{$S_n$}
\def\nt{\tilde n}
\def\izl{\int_0^L dx}
\def\Ad{\dot A_1}
\def\Wd{\dot W}

\def\de{\delta}
\def\eps{\epsilon}
\def\cc{conjugacy class}
\def\un{{\rm un}}
\def\ws{world-sheet}
\def\QCD{${\rm QCD}_2$}

\def\ts{target space}
\def\td{two-dimensional}

\def\pf{{g^2L\over2N}}
\def\ad{a^\dagger}

\def\NG{Nambu-Goto}

\Pubnum={CERN-TH-6843/93\cr
UVA-HET-93-02\cr
hepth@xxx/9303153}
\date={March 1993}
\pubtype={}
\titlepage
\title{Equivalence of Two Dimensional QCD and the $c=1$ Matrix Model}
\bigskip
\author {
Joseph~A.~Minahan\footnote\star
{minahan@gomez.phys.virginia.edu}}
\address{Department of Physics,  Jesse Beams Laboratory,\break
University of Virginia, Charlottesville, VA 22901 USA}
\andauthor{
Alexios P. Polychronakos\footnote\dagger
{poly@dxcern.cern.ch}}
\address{Theory Division, CERN\break
CH-1211, Geneva 23, Switzerland}
\bigskip
\abstract{
We consider two dimensional QCD with the spatial dimension compactified to a
circle. We show that the states in the theory consist of interacting strings
that wind around the circle and derive the Hamiltonian for this theory in the
large $N$ limit, complete with interactions.  Mapping the winding states into
momentum states, we express this Hamiltonian in terms of a continuous field.
For a $U(N)$ gauge group with a background source of Wilson loops, we recover
the collective field Hamiltonian found by Das and Jevicki for the $c=1$ matrix
model, except the spatial coordinate is on a circle. We then proceed to show
that two dimensional QCD with a $U(N)$ gauge group can be reduced to a
one-dimensional unitary matrix model and is hence equivalent to a theory of $N$
free nonrelativistic fermions on a circle. A similar result is true for the
group $SU(N)$, but the fermions must be modded out by the center of mass
coordinate.
}
\vfill
\endpage

\def\NP{{\it Nucl. Phys.\ }}
\def\PL{{\it Phys. Lett.\ }}
\def\PRD{{\it Phys. Rev. D\ }}

\def\PRL{{\it Phys. Rev. Lett.\ }}

\def\IJMP{{\it Int. Jour. Mod. Phys. A\ }}
\def\MPL{{\it Mod. Phys. Lett. A\ }}

\def\ZETF{{\it Zh. Eksp. Teor. Fiz.}}

\REF\tHooft{G.~'t~Hooft, \NP {\bf B75} (1974) 461.}
\REF\BBHP{W.~Bardeen, I.~Bars, A.~Hanson and R.~Peccei, \PRD {\bf13} (1976)
2364.}
\REF\BarsI{I.~Bars, \PRL {\bf36} (1976) 1521; \NP {\bf B111} (1976) 413.}
\REF\Gross{D.~Gross, LBL and Princeton preprints LBL 33233, PUPT 1356,
hepth/9212249; LBL 33232, PUPT 1355, hepth/9212248, 1992.}
\REF\JM{J.~Minahan, Virginia preprint, UVA-HET-92-10, hepth/9301003, to be
published in {\it Phys. Rev. D.}}
\REF\GrTay{D.~Gross and W.~Taylor, LBL-33458 hepth/9301068.}
\REF\GrTayII{D.~Gross and W.~Taylor, CERN-TH-6827-93 hep-th/9303046.}
\REF\Dalley{S.~Dalley, \MPL {\bf 7} (1992) 1651.}
\REF\Migdal{A.~Migdal, \ZETF {\bf 69} (1975) 810.}
\REF\Rus{B.~Rusakov, \MPL {\bf5} (1990) 693.}
\REF\Sakita{A.~Jevicki and B.~Sakita, \NP {\bf 165} (1980) 510;
B.~Sakita, {\it Quantum Theory of Many-Variable Systems and Fields},
World Scientific, 1985}
\REF\DJ{S.~Das and A.~Jevicki, \MPL {\bf 5} (1990) 1639.}
\REF\KS{D.~Karabali and B.~Sakita, \IJMP {\bf6} (1991) 5079.}
\REF\Demet{K.~Demeterfi, A.~Jevicki and J.~Rodrigues, \NP {\bf B362} (1991)
173.}
\REF\APA{A.P.~Polychronakos, \PL {\bf B266} (1991) 29.}
\REF\APG{A.P.~Polychronakos, \PL {\bf B264} (1991) 362.}

\chapter{Introduction}

Two dimensional QCD (\QCD) might prove to be a useful laboratory for exploring
some properties of the confining phase of four dimensional QCD.  This program
was originally started by 't Hooft[\tHooft] who computed the meson
spectrum in the planar limit.  He showed that asymptotically the states
live on a Regge trajectory.  Other researchers later demonstrated
that this spectrum could be derived from a \NG\ action[\BBHP,\BarsI].

More recently it was postulated that \QCD\ could be interpreted as a
theory of maps of two dimensional \ws s into a two dimensional \ts[\Gross].
This investigation was carried out further in [\JM] where it was shown
that the free energy of \QCD\ was consistent with a sum over maps containing
tubes and handles and it was also shown to low orders that the
counting of branched surfaces was consistent.  Finally in [\GrTay] it was
proven completely that \QCD\ is described by a sum over branched maps with
tubes into any two-dimensional \ts, except for some anomalous terms that
appear for \ts s with genus greater than one[\GrTayII].

Since \QCD\ is a string theory, it is natural to ask how it
compares with another well known \td\ string theory, the $c=1$ matrix model.
In [\Dalley] it was shown that the Weingarten model in two dimensions does
indeed lead to this model, if the spatial dimension is compactified onto
a vanishingly small circle.  Hence by comparing \QCD\ with $c=1$ matrix
models we are also
indirectly comparing it with the Weingarten model.

In this paper we consider $SU(N)$ and $U(N)$ \QCD\ on a
cylinder with circumference $L$.
In section two we construct the states of this system and argue that they
are described by strings that wrap around the compactified dimension.
The theory does not contain zero winding excitations.  Using the rules
developed in [\GrTay] for \QCD\ string theory, we derive
the complete Hamiltonian for the theory which describes strings joining
or breaking apart and also contains a potential term that describes
a ferromagnetic-like interaction between the strings.  This last term is
absent in the $U(N)$ case.  In section three
we map the winding states to momentum states on a circle with circumference
$4\pi/g^2L$, where $g/\sqrt{N}$ is the QCD coupling. At the large-$N$ limit
we show that the Hamiltonian reduces to the Das-Jevicki
Hamiltonian for a collective coordinate field, with the spatial coordinate
living on the circle.  The lack of zero winding
excitations is important in this derivation.  To reach the critical
$c=1$ theory, it is necessary to turn on a background source of Wilson loops.
In section four we show that \QCD\ on a cylinder reduces to the singlet sector
of a one-dimensional
unitary matrix model, the unitary matrix being the monodromy of the gauge
field around the compact space dimension. Therefore, the spectrum can be
reproduced by a theory of $N$ nonrelativistic free fermions on a circle.
This gives a natural explanation to the appearance of the collective field
Hamiltonian as well as to the origin of nonperturbative corrections.
If the gauge group is $SU(N)$ then the theory is
modded out by the center of mass coordinate.  In the
final section we present our conclusions.

\chapter{Derivation of the Hamiltonian}

Consider $SU(N)$ \QCD\ living on a torus with area $A$.  Its partition
function is given by[\Migdal,\Rus]
$$Z=\sum_{\rm reps}\exp(-Ag^2C_{2R}/N),\eqn\partfun$$
where the sum is over all representations of $SU(N)$,
$g/\sqrt{N}$ is the QCD coupling, and $\CR$ is the quadratic
Casimir of the representation.
A given representation $R$ is associated with a Young tableau
described by $m$ rows, with $n_i$ boxes in row $i$, which satisfy $n_i\ge n_j$
if $i<j$.  $\CR$ is then given by
$$\CR={N\over2}(n+{\nt\over N}-{n^2\over N^2}),\eqn\quadcas$$
where
$$n=\sum_{i=1}^mn_i,\qquad\qquad\nt=\sum_{i=1}^mn_i(n_i-2i+1).\eqn\ntndef$$

Let us describe the torus by two circles with circumferences $\be$ and
$L$ so that $A=\be L$. $\be$ can be though of as the inverse temperature,
therefore,
the partition function describes \QCD\ at finite temperature with its  spatial
dimension compactified onto a circle  of length $L$.  From \partfun\ it is
clear that every representation of $SU(N)$
corresponds to a physical state of the theory,
with energy $(g^2L/N)\CR$.  A state in representation $R$ is created and
destroyed by a Wilson loop
in representation $R$ that wraps once around the spatial dimension.
To see this, we can consider a cylindrical surface with Euclidean length
$\be$ and Wilson loops with representations $R$ and $R'$
inserted at the two ends of the cylinder.  This partition function is given
by
$$\eqalign{Z&=\int d\Om d\Om'\chi_R(\Om)\chi_{R'}(\Om'^{-1})
\sum_{R''}\chi_{R''}(\Om) \chi_{R''}(\Om'^{-1})\exp(-g^2\be L C_{2R''})\cr
&=\de_{RR'}\exp(-g^2\be L\CR),}\eqn\partcyl$$
where $\Om$ and $\Om'$ are the $SU(N)$ elements around the circles at the ends
of the cylinder and $\chi_R(\Om)$ and $\chi_{R'}(\Om'^{-1})$ are the
corresponding characters.
Clearly, the partition function in \partcyl\ represents the
propagation of one state into itself over a euclidean time $\be$.

Recently it was shown that \QCD\ has a string theory
interpretation[\Gross-\GrTay].  That is,
the partition function can be thought of as a set of maps of two-dimensional
\ws s into a two-dimensional \ts.  The maps can multiply cover the surface,
and such maps can contain branch cuts or small tubes that connect the different
sheets of the \ws.

For a given representation $R$,
the expression $\exp[-(g^2A/2)(n+\nt/N-n^2/N^2)]$
can be expanded in powers of $1/N$.  The leading term is $\exp(-g^2An/2)$,
hence this representation describes an $n$-covered map, with the leading
term coming from the integration of the \NG\ action over the \ws.  The
expansion of $\exp[-(g^2A/2)(\nt/N)]$ is the contribution of the branch
cuts connecting the sheets and the expansion of $\exp[-(g^2A/2)(-n^2/N^2)]$
gives the contributions of the tubes and small handles.

We also must consider the complex conjugate representations of $R$, $\bR$.
One can think of the sheets for this representation as having the opposite
chirality to those of representation $R$.  It is also possible
to have representations which are tensor products of $R$ and $\bR'$.  For
such representations there are no branch points connecting sheets of opposite
chirality and tubes that connect such sheets come with a minus sign[\GrTay].

Now consider the string picture for the cylinder with \Wl s inserted at
the ends.  A chiral representation $R$, with $n$ boxes in its tableau is a
linear combination of string states that wrap around the compact dimension
a total of $n$ times, all in the same direction.  Hence, there could be
$n$ strings that wrap once, or one string that wraps around $n$ times.  The
total number of such states is $P(n)$, the number of partitions of $n$.

The branch points on the \ws\ correspond to interactions where two strings
join to form one string or {\it vice versa}.  The tubes correspond to
interactions where two strings ``kiss'' at a point
and break apart again, or a multiwound
string which bumps into itself.  Since there are no branch points joining
sheets of opposite chirality, two strings of opposite winding will not join
to form a single string,
nor will a string break into strings with opposite winding.
However, two states with opposite winding can have a pointlike interaction,
but the sign is opposite to that of two strings with the same winding.

The almost triviality of the interactions for two strings with opposite
winding essentially allows us to separate the two sectors.
With this in mind, consider a state with $n$ windings in one direction.
Following the work of Gross and Taylor[\GrTay], a string state can be described
by an
element of the permutation group for $n$ elements, \Sn.  At the spatial point
$x=0$, a label can be assigned to each of the $n$ strands of string.
Tracing the strands form $x=0$ to $x=L$, we find that some strands come
back to themselves, but others are mapped to different strands.
This mapping is described an element $s$, of \Sn.
So for example, the state with $n$ strings
that wind once is given by the identity element.  For any $t$ in \Sn, the state
$tst^{-1}$ corresponds to a relabeling of the strands, hence this state is
equivalent to the state described by $s$.  Therefore, the inequivalent states
are given by the conjugacy classes of \Sn.

We can define an inner product
$$\langle s'|s\rangle=\de_{s's},$$
where $s$ and $s'$ are elements of \Sn, but it is more useful to define the
unnormalized product
$$\eqalign{\langle s'|s\rangle_{\un}=\sum_{t\in S_n}\de_{s',tst^{-1}}
&={n!\over C_s}\qquad\qquad{\rm if}\ s\simeq s'\cr
&=0\qquad\qquad {\rm otherwise.}}\eqn\innerprod$$
The symbol $\simeq$ means that the elements are equivalent up to a conjugacy
and $C_s$ is the number of elements in the \cc.
Each \cc\ is described by a partition of $n$, where each element of the
class is a cycle within the elements of the partition.  If $s=s'$, then
the elements of \Sn\ which commute with $s$ are those elements which are
cycles of $s$, multiplied by those elements which exchange cycles with equal
number of elements.  If the partition is given by
$$\prod_{l=1}^n(l)^{nl},\qquad\qquad\sum_{l=1}^n ln_l=n,$$
then the order of the subgroup that commutes with $s$ is
$$\prod_{l=1}^n(l)^{nl}n_l!.$$
Hence this particular state can be written as
$$\prod_{l=1}^n(\ad_l)^{n_l}|0\rangle,\eqn\state$$
where $\ad_l$ is the creation operator for a string with winding $l$ and
$|0\rangle$ is the vacuum state.  We can also act on the vacuum with the
operators $\ad_{-l}$ which are the creation operators for strings that wind
in the opposite direction.  The commutation relations are given by
$$[a_l,\ad_m]=|l|\de_{l,m},\eqn\commrel$$
thus the inner products of these states will reproduce the result in
\innerprod.

To leading order in $1/N$, the energy of such a state is given by
$g^2L(n_l+n_r)/2$, where $n_l$ is the number of left windings and $n_r$ is
the number of right windings.
Hence, the leading order Hamiltonian is given by
$$H_0={g^2L\over2}\sum_{n\ne0}\ad_na_n.\eqn\Hamfree$$

Now consider the interactions among the strings.  At a branch point two strings
join or break apart.  As far as the permutations of the strands are concerned,
this corresponds to inserting an element of \Sn\ which has one cycle of order
2 and $n-2$ cycles of order 1.  One should
sum over all possible branch points, which corresponds to summing over
the entire \cc\ of these elements.  Therefore, the unnormalized matrix
element describing this interaction is given by
$$\sum_{p\in S_{n2}}\langle s'|p|s\rangle_\un=\sum_{t\in S_n\atop p\in S_{n2}}
\de_{s'p,tst^{-1}},\eqn\insertp$$
where $S_{n2}$ are the elements of $S_n$ in the \cc\ with one 2-cycle and the
rest 1-cycles.
If $s$ is comprised of two cycles of order $n_1$ and $n_2$ and $s'$ is
comprised of one cycle of
order $n_1+n_2$, then there is a unique $p$ such that
$s'p=s$.  Consider the set of elements in \Sn\ which are given by $t=rq$,
where $qsq^{-1}=s$ and $r^{-1}s'r=s'$.
The elements $q$ form a subgroup of order $n_1n_2$, while the elements $r$ form
a subgroup of order $n_1+n_2$.
Moreover, the conjugates of $p$, $r^{-1}pr$ form $n_1+n_2$ distinct
elements.
Hence the sum in \insertp\ is given by $(n_1+n_2)n_1n_2$.
$s$ and
$s'$ could also have additional cycles, but these are basically spectators
as far as $p$ is concerned, so the matrix elements for these states
can be determined using \innerprod.
Since each branch point comes with a factor $g^2/2N$,  and since
the branch point can occur anywhere along the circle of length $L$,
then in terms of the creation and
annihilation operators, the operator that leads to the matrix element in
\insertp\ is
$$\pf\ad_{n_1+n_2}a_{n_1}a_{n_2}.\eqn\branchint$$
Including windings in both
sectors, one then finds that the general Hamiltonian describing this class of
interactions is given by
$$H_b=\pf\left(\sum_{n,n'>0}+\sum_{n,n'<0}\right)
(\ad_{n+n'}a_{n}a_{n'}+{\rm c.c.}).\eqn\Hambranch$$

Finally, it is easy to see that the interaction term that describes
the handles and tubes on the \ws\ is given by
$$H_t={g^2L\over2N^2}\left[\sum_{n>0}(\ad_na_n-\ad_{-n}a_{-n})\right]^2.
\eqn\Hamtube$$
The operator inside the square brackets counts the net winding number
of the state.

So far in this section we have been assuming that the gauge group is
$SU(N)$ instead of $U(N)$.
One disadvantage of $SU(N)$ is that the winding number is actually
only defined modulo $N$.  That is, a state that has one string with
winding number $-1$ is
equivalent to a linear combination of strings with winding number $N-1$.
If the group is enlarged to $U(N)$, then this is no longer the case.  For a
$U(N)$ representation the quadratic Casimir has the additional term
${1\over2}n^2(g'^2/g^2)$, where $g'$ is the $U(1)$ coupling, $g$ is the
$SU(N)$ coupling and $n$ is the total net winding of left and right string
states.  Hence by considering $U(N)$, the interaction strength of the term in
\Hamtube\ becomes an adjustable parameter, and for $g'=g/N$, it can be
eliminated entirely.

\chapter{Derivation of the Das-Jevicki Hamiltonian}

A striking feature of the Hamiltonian given in \Hamfree, \Hambranch\
and \Hamtube\ is that the creation
and annihilation operators for the winding states look just like operators
which create and destroy momentum states in one spatial dimension.
With this in mind, define a new length
$\tL=4\pi/(g^2L)$.  We can then define a momentum variable as $k=2\pi n/\tL$,
where $n$ is the winding number.  In order to consider $k$ as momentum
excitations in a continuous space, it must take arbitrarily large values;
therefore, $n$ must be
very large.  On the other hand, the interacting boson picture breaks down if
$n\ge N$, since in this case some states end up being
summed over that do not correspond to representations of $U(N)$ or $SU(N)$.
Thus, the continuum limit is only valid in the large-$N$ limit.

Letting $a_k=a_n$, the commutation relation
in \commrel\ becomes
$$[a_k,\ad_{k'}]={\tL \over 2\pi}|k|\de_{k,k'}.\eqn\commrelk$$
We can also define a field $\vp(x)$ and its canonical
conjugate field $\Pi(x)$, where  $[\vp(x),\Pi(y)]=i\de(x-y)$.
We can then write $a_k$ as
$$\eqalign{
a_k&=\half\int dx e^{-ikx}[\vp(x)+\opi\pPi(x)],\qquad\qquad k>0\cr
&=\half\int dx e^{-ikx}[\vp(x)-\opi\pPi(x)],\qquad\qquad k<0
}\eqn\akeq$$
which one can easily show satisfies the commutation relations.

We now plug these expressions into the full Hamiltonian given in
\Hamfree, \Hambranch\ and \Hamtube.  First substituting $k$ for $n$,
we find that the complete Hamiltonian is given by
$$\eqalign{H={2\pi\over\tL}\sum_k & \ad_k a_k
+{2\pi\over\tL N}\left\{
\sum_{k,k' >0}(\ad_{k+k'}a_ka_{k'} +{\rm c.c.})
+\sum_{k,k' <0}(\ad_{k+k'}a_ka_{k'} +{\rm c.c.})
\right\}\cr
&-{2\pi\al\over\tL N^2}\left(\sum_{k>0} (\ad_ka_k-\ad_{-k}a_{-k})\right)^2
,}\eqn\Hamfull$$
where $\al$ is an adjustable parameter which depends on the $U(1)$ coupling.
Substituting the expression for $a_k$ in \akeq\ and performing the
sums over momenta then gives
$$\eqalign{H=&{1\over2\pi}\int dx(\pi^2\vp^2+(\pPi)^2)
+{\tL\over4\pi N}\int dx(\pi^2\vp^3+3\pPi\vp\pPi)\cr
&-{1\over4\pi\tL N}\Biggl\{\int dx\pi^2\vp(x)\left[\int dy\vp(y)\ct{x-y}
\right]^2\cr
&\qquad\qquad+\int dx\left[\int dy\pPi(y)\ct{x-y}\right]\vp(x)
\left[\int dz{\pPi(z)\ct{x-z}}\right]\cr
&\qquad\qquad
+2\int dx\pPi(x)\int dy\vp(y)\ct{x-y}\int dz\pPi(z)\ct{x-z}
\Biggr\}\cr
&-{\tL\al\over 2\pi N^2}\left[\int dx \vp\pPi\right]^2
+\Delta H,}\eqn\Hamx$$
where $\Delta H$ is the singular term,
$$\Delta H={\tL\over4\pi N}\int dxdy\de(x-y)\vp(x)\partial_x\partial_y
\ln|\sin{\pi \over \tL}(x-y)|.\eqn\Hcorr$$

Let us now set $\al$ to zero.  The Hamiltonian we are left with is still
non-local,  but there is an important property of \QCD\ which will improve
this situation.  \QCD\
contains no zero winding excitations and in fact, no
such terms appear in \Hamfree, \Hambranch\ or \Hamtube.  Therefore,
$\vp(x)$ and $\pPi(x)$ can not contain zero modes.  Thus, one
must impose the constraints
$$\int dx \vp(x)=\int dx\pPi(x)=0.\eqn\constraint$$

Since the integrals in \constraint\ are finite, the non-local pieces can
be expressed in terms of local terms by using a somewhat modified trick of
collective coordinate field theories[\Sakita].
Defining $\tf(x)$ as
$$\tf(x)={\pi\over\tL}\int dx \ct{x-y}f(y),\eqn\tfeq$$
we have that
$${\pi\over\tL}\int dyf(y)\ct{x-y\pm i\eps}=\tf(x)\mp i\pi f(x),\eqn\sintrans$$
where $f(x)$ needs to be reasonably smooth and $\tf(x)$ must exist.
Taking the identity
$$\eqalign{\ct{x-y+i\eps}\ct{y-z+i\eps}&+\ct{y-z+i\eps}\ct{z-x-2i\eps}\cr
&+\ct{z-x-2i\eps}\ct{x-y+i\eps}=1,}\eqn\cotrel$$
multiplying it by $f(x)g(y)h(z)$, and then integrating over $x$, $y$, and $z$,
we find that
$$\int dx fgh=\int dx [f\tg\th+\tf g\th+\tf\tg h]
+{\pi^2\over\tL^2}\int dxf(x)\int dyg(y)\int dz h(z).\eqn\Hilbrel$$
Using \Hilbrel\ and the constraint \constraint,
we are now able to rewrite the Hamiltonian in \Hamx\ as
$$H={1\over2\pi}\int dx\left\{\pi^2\vp^2+(\pPi)^2+{\tL\over N}
\left[{\pi^2\over3}\vp^3+\pPi\vp\pPi\right]\right\}
+\Delta H.\eqn\Hamagain$$
Shifting $\vp$ to $\vp+N/\tL$, the Hamiltonian becomes
$$\eqalign{H&={\tL\over2\pi N}\int dx\left\{\pPi\vp\pPi+{\pi^2\over3}\vp^3
-\left({\pi N\over\tL}\right)^2\vp\right\}+\Delta H\cr
&= {4\over g^2LN}\int dx\left\{\half\pPi\vp\pPi+{\pi^2\over6}\vp^3
-\left({g^2LN\over4}\right)^2\vp\right\}+\Delta H,}\eqn\HamDJ$$
up to a constant.  Moreover, the new constraint becomes
$$\int dx\vp(x)=\tL N/\tL=N.\eqn\conDJ$$

Except for a missing potential term and the fact that the fields live on a
circle as opposed to in a box,
the Hamiltonian in \HamDJ\ and the constraint in \conDJ\ are precisely
those found by Das and Jevicki for the collective coordinate field of
the $c=1$ matrix model[\DJ].  $\Delta H$ is the quantum correction to the free
energy[\DJ,\KS].
This calculation is also analogous to one in [\Demet], but with
different boundary conditions.
 From \HamDJ\ we see that the bare string coupling
constant is $4/(g^2LN)$, thus strong coupling QCD leads to a weak coupling
string theory.  The constraint in \conDJ\ can be imposed by adding the term
$$\int dx \muF(\vp-{N\over\tL})$$
to the Hamiltonian, where $\muF$ is a Lagrange multiplier which acts as the
bare cosmological constant.  Of course, the linear term in \HamDJ\ will shift
this value.

To complete the program, we need to have a potential term,
$\int dx V(x)\vp(x)$ in the Hamiltonian, so that theory can have some sort
of critical behavior.  Such a term in momentum space is given by
$$\sum_k V_k(a_k+\ad_k).\eqn\potmom$$
Such terms can be produced by a background source of \Wl s.  For instance,
if $V(x)=\cos{\pi \over \tL} x$, then $V_k=a_1+\ad_1+a_{-1}+\ad_{-1}$.
Hence the \QCD\ action should contain the additional term
$$C\int dt[\chi_f(U(t))+\chi_{\bar f}(U(t))],\eqn\wleq$$
where $U(t)$ is the value of the $U(N)$ element around a closed loop at time
$t$ and $\chi_f(U)$ and $\chi_{\bar f}(U)$ are the characters for the
fundamental representation  and its complex conjugate.  Note that
$\chi_f(U)$ creates strings that wind to the right and annihilates strings
that wind to the left.  Critical behavior can now be found by tuning $C$.
This has the same perturbative behavior as the $c=1$ Hermitian matrix model,
but its nonperturbative behavior is different because of the different
boundary conditions.

\chapter{Free Fermions}

The fact that a collective coordinate field theory can be constructed
using the rules derived from \QCD\ string theory suggests that there
exists a free fermion picture of \QCD.
In this section we show that \QCD\ on a cylinder {\it is} equivalent to a
theory of free fermions by showing that it can be reduced to a one-dimensional
unitary matrix model.

To this end, consider \QCD\ in the gauge $A_0=0$.
The Hamiltonian is then given as
$$H=\half\izl\tr F_{01}^2=\half\izl\tr \Ad^2\eqn\Hamgauge$$
with the overdot denoting a time derivative.
The $A_0$ equation of motion is now the constraint
$$D_1F_{10}=\partial_1\Ad+ig[A_1,\Ad]=0.\eqn\gconstraint$$
Let us now define a new variable $V(x)$,
$$V(x)=W_0^x\Ad(x)W_x^L,\eqn\Vdef$$
where
$$W_a^b={\rm P}e^{ig\int_a^bdxA_1}.\eqn\Wdef$$
Then \Hamgauge\ can be written as
$$\partial_1V(x)=0,\eqn\Veq$$
so $V(x)$ is a constant.  Thus $V(0)=V(L)$, which implies that
$$[W,\Ad(0)]=0,\eqn\WAcomm$$
where $W\equiv W_0^L$ and we have used the periodicity of $A_1$ in $x$.

 From the definitions \Vdef\ and \Wdef, we find the relation
$$\Wd=ig \izl W_0^x\Ad(x)W_x^L=ig \izl V(x),\eqn\Wdeq$$
and therefore using \Veq\ and \WAcomm, we derive
$$\Wd=igLW\Ad(0)=igL\Ad(0)W.\eqn\WdeqII$$
\WdeqII\ then implies that
$$[W,\Wd]=0.\eqn\WWdeq$$

Because $V(x)=V(0)$, $\Ad(x)$ satisfies
$$\Ad(x)=W_0^x\Ad(0)W_x^0.\eqn\Adeq$$
Thus, using this relation along with \WdeqII, we can rewrite the Hamiltonian
in \Hamgauge\ as
$$H=-{1\over 2g^2L}\tr(W^{-1}\Wd)^2.\eqn\Hammm$$
If the gauge group is $U(N)$, with the $U(1)$ coupling given by $g/N$, then
\Hammm\ is the Hamiltonian for the one-dimensional unitary matrix model.
The canonical structure of this Hamiltonian is also the standard matrix
model one, as can be deduced from the fundamental brackets
$$ \{ A_1 (x)_{ij} , {\dot A}_1 (y)_{kl} \} = \de_{il}\,
\de_{jk} \, \de(x-y) \eqn\PB$$
and the definition of $W$.
The constraint in \WWdeq\ reduces the space of states to singlets[\APA].
Hence, the problem is reducible to the eigenvalues of $W$.

Upon quantization, this problem is equivalent to a system of $N$
nonrelativistic fermions living on a circle, with the Hamiltonian given by
$$H=-\left({g^2L\over2}\right)\sum_{i=1}^N{\partial^2\over\partial\theta_i^2},
\qquad\qquad 0\le\theta_i<2\pi.\eqn\Hamferm$$
The fermionization is achieved by the appearance of the Vandermonde
determinant in the wavefunction of the states, which in the unitary
matrix case reads
$$ \Delta = \prod_{i<j} \sin{\theta_i - \theta_j \over 2}.\eqn\VM$$
Notice that each factor in \VM\ is antiperiodic on the circle.
Thus, if $N$ is even the fermions have antiperiodic boundary conditions.
Likewise, if $N$ is odd they have periodic boundary conditions.
This can be understood in terms of transporting a fermion once around the
circle, passing by $N-1$ other fermions along the way
and therefore picking up $N-1$ minus signs.
Hence, in either case, the ground state is built by filling all states
with wave numbers between $-N/2+1/2$ and $N/2-1/2$, inclusive.
Subtracting off the ground state energy, one easily sees that this spectrum
reproduces that found for the different representations of $U(N)$.

If the gauge group is $SU(N)$, because $A_1$ is now traceless
$W$ will also obey the condition $\det W = 1$. Therefore the center of
mass coordinate for the fermions is absent and we must mod it out
of the theory. This means that we need to identify states in which all
fermions have their momentum shifted by the same amount. (This is equivalent to
identifying the antisymmetric tensor product of $N$ copies of fundamental
representations with the singlet representation.)
Moreover, we must subtract the energy of the center of mass from the
energy of each state in the theory.

The correspondence of the fermion states with the Young tableaux is as
follows: since the center of mass coordinate drops out, we can always set
the smallest wave number to zero (for odd $N$.) The rest of the wave numbers
are integer numbers greater than zero, with the largest number equal to
$N-1$ for the ground state.  We can excite states by shifting the wave numbers
up (except the smallest one).   The size of the shift for the largest number
gives the number of boxes in the first row of the tableau, the size of the
next largest is the number of boxes in the second row, {\it etc.}
If we denote the shift of the $i^{\rm th}$ highest wave number as $n_i$,
then the energy of the state minus the ground state energy and the center
of mass energy is
$$\eqalign{E&={g^2L\over2}\Biggl\{\sum_i\Bigl[(n_i+N-i)^2-(N-i)^2\Bigr]\cr
&\qquad\qquad\qquad-
{1\over N}\left[\left({N(N-1)\over2}+\sum_i n_i\right)^2-\left({N(N-1)\over2}
\right)^2\right]\Biggr\}\cr
&={g^2L\over2}\left\{N\sum_in_i+\sum_in_i(n_i-2i+1)-{1\over N}
\left(\sum_in_i\right)^2
\right\}=g^2L\CR.}\eqn\sunenergy$$
After rescaling $g^2\to g^2/N$, we recover the expected result.
For even $N$ the argument is the same but with all the momenta shifted
by $1 \over 2$. Since this is a center of mass excitation, it does not
affect the energy and the same result is obtained.

In the string picture the two chiral sectors of the QCD partition
function are identified as excitations of left-moving or right-moving
fermions. The factorization of the two sectors
(that is, the fact that there are no states where a left-moving
fermion is excited into a right-moving state) holds up to leading order in
$1/N$ because the center of mass has been modded out.
This factorization, of course, completely
breaks down when a large number of quanta are excited (of order $N$)
which signals the onset of nonperturbative effects.

If the gauge group is $U(N)$ but with $U(1)$ coupling $g'\ne g/N$, then
the energies of the states are given by \Hamferm, but with a modified
coefficient for the center of mass kinetic energy operator.
Such a variable coefficient was discussed before in another context[\APG].

\chapter{Discussion}

We have found two main results.  The first is that the rules derived
from \QCD\ string theory lead to a collective coordinate theory
which is the same perturbatively as the collective coordinate theory
of $c=1$ matrix models. The second, which is related to the first, is that
\QCD\ is exactly equivalent to a theory of nonrelativistic fermions living on
the circle.  This theory differs from the usual $c=1$ theory nonperturbatively,
in that the fermions live on the circle instead of the real line.

The above results are quite encouraging and the extent to which they might
apply to higher dimensional QCD is an interesting issue.
Comparing the results of \QCD\ with those found for the Weingarten model,
shows a qualitative difference between the two models.
Unlike the Weingarten model it is not necessary to shrink $L\to0$ to
reach a $c=1$ matrix model.  The two theories also possess different
nonperturbative behavior.
Perhaps this will have some implications for higher dimensions.

These results might have more significance in understanding the anomalous
terms in the free energy for \QCD\ on a higher genus target space.
These terms can be reproduced by inserting special operators on the
\ws\ surface[\GrTayII], but their geometrical significance is yet to
be understood.

\ack{The research of J.A.M. was supported in part by D.O.E. grant
DE-AS05-85ER-40518.}
\refout
\end